\documentclass[]{spie}  

\usepackage{amsmath,amsfonts,amssymb}
\usepackage{graphicx}
\usepackage[colorlinks=true, allcolors=blue]{hyperref}
\usepackage{physics}

\title{Quantum tomography for quantum systems optimization}

\author{Boris I. Bantysh}
\author{Yurii I. Bogdanov}
\affil{Valiev Institute of Physics and Technology of Russian Academy of Sciences}

\authorinfo{Corresponding author e-mail: bbantysh60000@gmail.com}

\begin{document} 
\maketitle

\begin{abstract}
Debugging quantum states transformations is an important task of modern quantum computing. The use of quantum tomography for these purposes significantly expands the range of possibilities. However, the presence of preparation and measurement errors complicates the practical use of this procedure. In this work, we investigate the possibility of estimating these errors from experiment. These estimates are subsequently used to build a robust quantum tomography model. The model allows one to accurately reconstruct unitary errors of single-qubit gates. We show that, having such imperfect single-qubit gates with pre-estimated errors, one can obtain transformations close to ideal ones. A similar approach can also significantly mitigate single-qubit gates cross-talk.
\end{abstract}


\section{INTRODUCTION}

Over the past few years, a number of scientific groups have demonstrated impressive results in building the registers of qubits based on various physical principles: superconducting circuits \cite{arute2019quantum,jurcevic2021demonstration}, neutral atoms in optical traps \cite{bernien2017probing}, linear optical interferometers \cite{zhong2021phase}, and trapped ions \cite{zhang2017observation,wright2019benchmarking}. The latter is of particular interest, as trapped ions, in principle, make it possible to obtain high coherence times and high accuracy of one- and two-qubit gates \cite{ladd2010quantum,bruzewicz2019trapped,low2020practical}. In practice, however, experimental limitations still not allow achieving sufficiently high fidelity values of quantum transformations for chains of several ions.

Debugging of quantum transformations is often performed by optimization of control parameters with respect to randomized benchmarking curves \cite{muhonen2015quantifying,kelly2014optimal}. This approach requires a complex optimization procedure and the collection of a large amount of experimental data. This is mainly due to the fact that each individual experiment outputs only a single number characterizing the quantum gate fidelity. At the same time, quantum tomography (QT) allows one to obtain a complete description of a quantum transformation and determine how it affects any given input state \cite{knee2018quantum,teo2020objective,bialczak2010quantum,bogdanov2013quantum,rodionov2014compressed,bantysh2019high}.

However, at the moment, there are practically no papers where QT is used to optimize quantum transformations. The main reason for this is that QT is sensitive to quantum state preparation and measurement (SPAM) errors. In such conditions the estimates of the quantum process parameters are biased relative to their true values. To avoid this, the QT model must take into account SPAM errors \cite{bantysh2019high,schwemmer2015systematic,ramadhani2021quantum,bantysh2020precise,bantysh2020quantum,hou2016error,lvovsky2004iterative,bogdanov2016quantum}. One way to achieve this is through the use of machine learning methods \cite{fastovets2019machine,palmieri2020experimental}. However, errors in the training data can lead to the overestimated accuracy of subsequent results. The gate set tomography (GST) approach allows one to estimate the parameters of several transformations at once and automatically take into account SPAM errors \cite{blume2017demonstration,nielsen2020probing,dehollain2016optimization}. The disadvantage of GST is an ambiguity of the results obtained.

In this paper, we construct an unambiguous QT model from a set of measurements. To do this, we reduce the number of estimated parameters, taking into account some physical features behind the quantum states of trapped ions (Section~\ref{sect:noisy_qt}). The resulting QT model potentially makes it possible to completely eliminate instrumental errors in data and obtain reliable estimates of quantum gates parameters. These estimates can be used to optimize control parameters. In Section~\ref{sect:gates_opt}, however, we look for improving logical operations on qubits using existing imperfect quantum transformations. Our approach has a number of significant limitations. In Section~\ref{sect:discussion}, we discuss these limitations and the ways of reducing them.

\section{NOISY QUANTUM TOMOGRAPHY}\label{sect:noisy_qt}

\subsection{Standard quantum process tomography}\label{sect:qpt}

Consider a quantum process over a $d$--dimensional system. Its tomography is based on implementation of a set of quantum circuits. Each time there is a quantum state $\rho_j$ from the set $\mathcal{P}=\{\rho_1,\dots,\rho_{m_P}\}$ at the process input. Output state is measured in some basis $M_\alpha=\{M_{\alpha,1}, M_{\alpha,2},\dots\}$ ($M_{\alpha,k}$ is the POVM operator for the $k$-th result) from the set $\mathcal{M}=\{M_1,\dots,M_{m_M}\}$. Measurement protocol consists of $m=m_P\cdot m_M$ quantum circuits. Here we consider the most common standard protocol \cite{bantysh2020quantum,mohseni2008quantum,baldwin2014quantum}. For a single qubit it consists of $m_P=4$ input states and $m_M=3$ measurement bases.

\begin{figure}[ht]
    \begin{center}
        \begin{tabular}{c}
            \includegraphics[width=.9\linewidth]{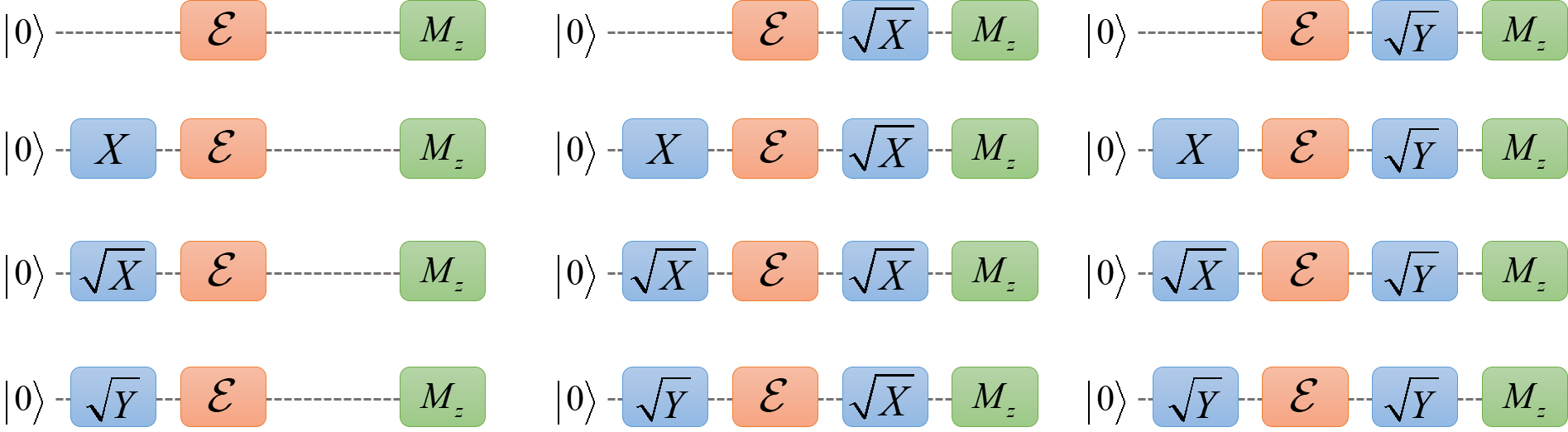}
        \end{tabular}
    \end{center}
    \caption{
        \label{fig:qt_circuits}
        The set of quantum circuits providing the standard protocol of single qubit process $\mathcal{E}$ tomography.
    }
\end{figure}

Usually to prepare an input state the system is initialized in the $\ketbra{0}$ state and transformed by some unitary channel. The measurement in some basis is done by the basis change operation and the readout in $z$-basis $\{\ketbra{0}, \ketbra{1}\}$. It turns out that for a single qubit one needs to perform only 4 different types of transformations: $I$ (identity transformation doing nothing to a qubit), $X=R_0(\pi)$, $\sqrt{X}=R_0(\pi/2)$, $\sqrt{Y}=R_{\pi/2}(\pi/2)$ (Figure~\ref{fig:qt_circuits}). Here $R_\varphi(\delta)=U(\pi/2,\varphi,\delta)$ and
\begin{equation}\label{eq:u_rotation}
    U(\theta,\varphi,\delta) = \begin{pmatrix}
        \cos{\frac{\delta}{2}}-i\cos{\theta}\sin{\frac{\delta}{2}} & -i\sin{\theta}e^{-i\varphi}\sin{\frac{\delta}{2}} \\[1em]
        -i\sin{\theta}e^{+i\varphi}\sin{\frac{\delta}{2}} & \cos{\frac{\delta}{2}}+i\cos{\theta}\sin{\frac{\delta}{2}}
    \end{pmatrix}
\end{equation}
is the transformation that rotates the state on the Bloch sphere ($\theta$ and $\varphi$ are the rotation axis spherical angles, $\delta$ is the rotation angle). For an optical trapped ion qubit $R_\varphi(\delta)$ is done by a single resonant laser pulse \cite{zalivako2021experimental}. Pulse phase determines $\varphi$, and its amplitude and duration determines $\delta$. A sequence of two pulses with different $\varphi$ and $\delta$ could give any single qubit rotation.

The data obtained from all $m$ measurements are used to reconstruct the parameters of the quantum process. There are various algorithms for performing this procedure \cite{knee2018quantum,teo2020objective,bialczak2010quantum,bogdanov2013quantum,rodionov2014compressed,bantysh2019high}. In this work, we are interested in the unitary part of the evolution of quantum states, since it is responsible for the logical operations on states. In this regard, we use the first rank root parameterization for the quantum process and the maximum likelihood estimation \cite{bogdanov2013quantum,bantysh2019high}.

Thus, in the case of a single qubit, a unitary quantum process can be described by three independent parameters (for example, the parameters from expression \eqref{eq:u_rotation}). In the general case, the number of independent parameters for a unitary process is $\nu=d^2-1$ \cite{bogdanov2013quantum}.

\subsection{Fuzzy measurements model}\label{sect:fuzzy_model}

Let $\rho_0$ be the true state the system is initialized in, and $\{P_0, \dots, P_{d-1}\}$ be the POVM operators of readout. Then for the circuit with $j$-th input and $\alpha$-th measurement
\begin{equation}\label{eq:fuzzy_operators}
    \rho_j = \mathcal{E}_j(\rho_0), \quad M_{\alpha,k}=\overline{\mathcal{E}}_\alpha(P_k), \quad k=0,\dots,d-1.
\end{equation}
Here $\mathcal{E}_j$ and $\mathcal{E}_\alpha$ are state preparation and basis change transformations respectively, and $\overline{\mathcal{E}}(P)=\sum_k{E_k^\dagger P E_k}$ is the map, conjugate to the process map $\mathcal{E}=\sum_k{E_k P E_k^\dagger}$. Even $\overline{\mathcal{E}}$ is generally not a proper quantum process, the set $\{M_{\alpha,1}, M_{\alpha,2},\dots\}$ forms POVM and gives non-orthogonal decomposition of the identity operator. Meanwhile, ideal projectors $\{\ketbra{0},\ketbra{1},\dots\}$ give orthogonal decomposition.

Operators \eqref{eq:fuzzy_operators} form fuzzy measurements model for QT. Using this model, it is possible to obtain a consistent estimate of a quantum state or a quantum process \cite{bantysh2019high,bantysh2020precise,bantysh2020quantum,bogdanov2016quantum}. Note, however, that this requires knowing the SPAM parameters with sufficient accuracy. Below we consider a possible experimental procedures for estimating these parameters in the trapped ion system. For simplicity, we assume the initialization being ideal ($\rho_0=\ketbra{0}$). This assumption is justified for trapped ion systems, where optical cooling gives precise initialization \cite{bruzewicz2019trapped}.

\subsection{Readout errors estimation}\label{sect:ro_estimation}

Trapped ion qubit readout is performed by detecting the fluorescence of the bright state for a certain finite period of time. Its duration must be high enough to reliably distinguish the bright state signal, taking into account the limitations of the detection system. At the same time, long-time readout is undesirable both by itself and as it increases the probability of the dark state to excite and fluoresce.

In practice, one usually introduce some threshold value for the number of registered photons \cite{zalivako2021experimental,christensen2020high}. If the readout shows the number of photons more than this threshold, then the result is ``0''. Otherwise, the result is ``1''. In this case, there is a nonzero probability $e_{10}$ to receive the number of photons less than the threshold (measurement result ``1'') even if the ion was in the bright state. There is also a probability $e_{01}$ that the ion in the dark state gives the number of registered photons above the threshold (measurement result ``0''). Thus, instead of the standard projective measurement with projectors $\ketbra{0}$ and $\ketbra{1}$, the measurement with the following POVM operators is implemented:
\begin{equation}\label{eq:fuzzy_readout}
    P_0=(1-e_{10})\ketbra{0} + e_{01}\ketbra{1}, \quad P_1=e_{10}\ketbra{0} + (1-e_{01})\ketbra{1}.
\end{equation}
Here we assume $\ket0$ ($\ket1$) to be the bright (dark) state.

The estimation of the optimal threshold value for the number of registered photons, as well as the parameters $e_{10}$ and $e_{01}$, can be performed experimentally. For this, the ion is initialized in the state $\ket0$ and measured. The ratio of the result ``1'' amount in a series of such experiments is the estimate of $e_{10}$. These experiments could be also repeated in the absence of one of the cooling lasers \cite{zalivako2021experimental}. This makes it possible to simulate the detector counts in the absence of fluorescence. The ratio of the result ``0'' amount in such experiments gives the estimate of $e_{01}$.

\subsection{QT gates estimation}\label{sect:qt_gates_estimation}

To perform QT of a single qubit operation one needs to implement gates $X$, $\sqrt{X}$, and $\sqrt{Y}$ (Section~\ref{sect:qpt}). Note, that $X$ is just $\sqrt{X}$ performed twice. To do $\sqrt{Y}$, one shifts the laser pulse phase by $\pi/2$. This phase appears in the transformation $\sqrt{Y}=Z(\pi/2)\cdot \sqrt{X}\cdot Z(-\pi/2)$, where $Z(\delta)=U(0,0,\delta)$. The error $e$ in the phase shift just shifts this transformation: $Z(\delta)\rightarrow Z(\delta+e)$.

\begin{figure}[ht]
    \begin{center}
        \begin{tabular}{ccc}
            \includegraphics[width=.3\linewidth]{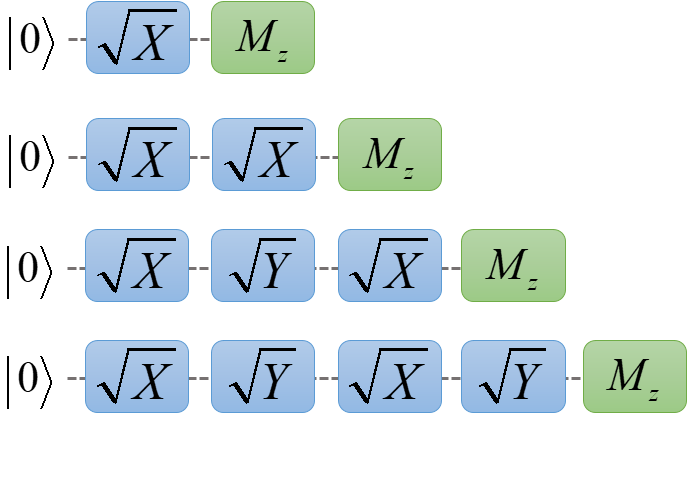} &
            &
            \includegraphics[width=.4\linewidth]{Fig2b.pdf} \\
            \textbf{a} & & \textbf{b}
        \end{tabular}
    \end{center}
    \caption{
        \label{fig:qt_model_circuits}
        (a) The set of quantum circuits, enough to reconstruct the QT gates.
        (b) QT gate reconstruction infidelity versus per circuit sample size $N$. Readout errors: $e_{10}=0.01$, $e_{01}=0.03$. Gates parameters error rate: $\varepsilon=0.01$. The median and the lower and upper quartiles for 100 independent numerical experiments for each $N$ are shown.
    }
\end{figure}

Thus, for our QT model we need to determine $\sqrt{X}$ parameters and $c=\pi/2+e$ only. Note, that the pulse absolute phase does not play a role, but only the relative pulse phase is important. Thus, we fix zero phase of gate $\sqrt{X}$ and parametrize it as $U(a,0,b)$. We then estimate three parameters only: $a$, $b$, and $c$. In the ideal noise-free case $a=b=c=\pi/2$, so one can restrict the search domain to $[0,\pi]$. Then four quantum circuits are enough to get unambiguous estimates of all parameters (Figure~\ref{fig:qt_model_circuits}a).

We have performed a number of numerical experiments. In each one random model parameters have been generated: $a=\pi/2+\varepsilon\cdot\mathcal{N}_a$, $b=\pi/2+\varepsilon\cdot\mathcal{N}_b$, $c=\pi/2+\varepsilon\cdot\mathcal{N}_c$, where $\mathcal{N}_{a,b,c}$ are independent standard normal random variables, and $\varepsilon$ is the error rate. The reconstruction has been done using least-squares method taking operators \eqref{eq:fuzzy_readout} into account. Gates accuracy have been estimated using the fidelity measure
\begin{equation}\label{eq:fidelity}
    F = \frac{1}{d^2}\abs{\Tr(U^\dagger V)}^2
\end{equation}
between the expected $U$ and true $V$ unitaries. The results are shown in Figure~\ref{fig:qt_model_circuits}b. One can observe that the infidelity decreases as $\propto1/N$ with increasing per circuit sample size $N$.

\section{GATES OPTIMIZATION}\label{sect:gates_opt}

\subsection{Rotation angles}\label{sect:gates_angles}

Transformation $R_\varphi(\delta)$ of a trapped ion qubit is determined by $\varphi$ and $\delta$, which depend on the control pulse form. If the pulse parameters are not well tuned the resulting transformation will be $\tilde{R}_\varphi(\delta)=U(\theta_r,\varphi_r,\delta_r)$. Let us assume that the dependence $(\theta_r,\varphi_r,\delta_r)$ of $(\varphi,\delta)$ has the linear form:
\begin{equation}\label{eq:linear_model}
    \begin{aligned}
        \theta_r(\varphi, \delta) &= a_{11}\varphi + a_{12}\delta + a_{13}, \\
        \varphi_r(\varphi, \delta) &= a_{21}\varphi + a_{22}\delta + a_{23}, \\
        \delta_r(\varphi, \delta) &= a_{31}\varphi + a_{32}\delta + a_{33}.
    \end{aligned}
\end{equation}
In the ideal case the matrix of coefficients is
\begin{equation}\label{eq:linear_model_ideal}
    \qty[a_{kl}^{id}] = \begin{pmatrix}
        0 & 0 & \pi/2 \\
        1 & 0 & 0 \\
        0 & 1 & 0
    \end{pmatrix}.
\end{equation}

Knowing the $\tilde{R}_\varphi(\delta)$ for any $\varphi$ and $\delta$ one can combine them to get the required transformation. If $a_{kl}$ differ from $a_{kl}^{id}$ by a few percents, it is enough to implement two-gate sequence $\tilde{R}_{\varphi_2}(\delta_2)\tilde{R}_{\varphi_1}(\delta_1)$ for any given single qubit rotation. For larger error rates it might take longer sequences.

Let us take a set of gates with different expected parameters $(\varphi,\delta)$. We perform the QT over each of them and extract the corresponding true values $(\theta_r,\varphi_r,\delta_r)$. We then find a join solution of \eqref{eq:linear_model} over $a_{kl}$ for all of these gates. This gives us the estimate $\hat{a}_{kl}$ of the model coefficients.

For simulation we have used the set of four transformations with parameters
\[
(\varphi,\delta)=\{(\pi/4,\pi/2),(\pi/4,\pi),(3\pi/4,\pi/2),(3\pi/4,\pi)\}.
\]
We have generated a random model coefficients as $a_{kl}=a_{kl}^{id}+\varepsilon\mathcal{N}_{kl}$, where $\mathcal{N}_{kl}$ are independent standard normal random variables, and $\varepsilon$ is the error rate. We have then performed the QT of the chosen gates and extracted the model coefficients estimates $\hat{a}_{kl}$. We have used them to optimize $\tilde{R}_{\varphi_2}(\delta_2)\tilde{R}_{\varphi_1}(\delta_1)$ over $(\varphi_1, \delta_1, \varphi_2, \delta_2)$ to get a desired rotation.

\begin{figure}[ht]
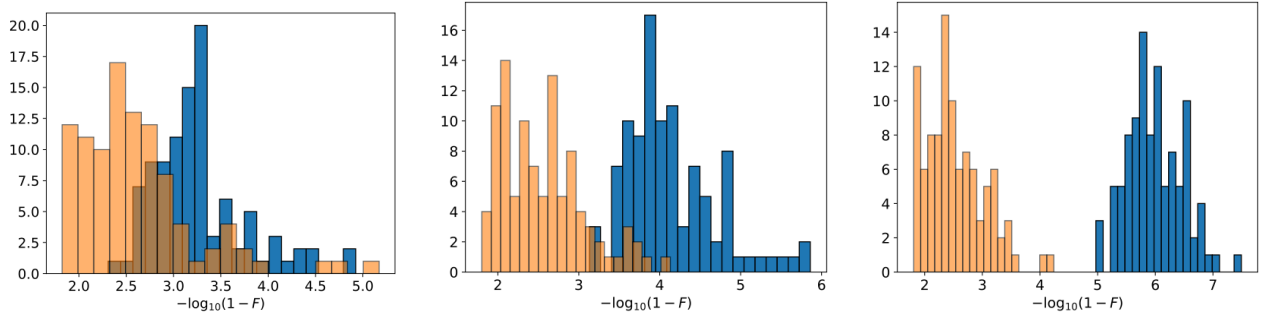

    \begin{center}
        \begin{tabular}{lcr}
            \includegraphics[width=.31\linewidth]{Fig3a.pdf} &
            \includegraphics[width=.31\linewidth]{Fig3b.pdf} &
            \includegraphics[width=.31\linewidth]{Fig3c.pdf}
        \end{tabular}
    \end{center}
    \caption{
        \label{fig:angles_result}
        Distributions of the accuracy of the single qubit transformation implementation using reconstructed (blue) and standard (orange) models in \eqref{eq:linear_model}. Error rate $\varepsilon=0.01$. A total of 100 numerical experiments have been performed. In each of them a transformation of the form \eqref{eq:u_rotation} with random parameters has been generated. To estimate $a_{kl}$ using QT the sample size of $10^3$ (left), $10^4$ (center), and $10^5$ (right) has been used for each quantum circuit.
    }
\end{figure}

Figure~\ref{fig:angles_result} shows the comparison of the obtained model with the standard one that considers \eqref{eq:linear_model_ideal}. The quantity $-\log_{10}(1-F)$ shows number of nines after decimal point in fidelity \eqref{eq:fidelity}. One can observe that the fidelity about 99.99\% needs about $10^4$ samples for every quantum circuit at the stage of coefficients $a_{kl}$ estimation.

\subsection{Single qubit gate cross-talk}\label{sect:gates_ct}

Experimental limitations do not always allow a precise addressing of single qubit transformations in the chain of several ions: the neighboring ions are also transformed. More accurate focusing and the introduction of a spatially inhomogeneous magnetic field can significantly suppress this effect \cite{bruzewicz2019trapped}. Alternatively, one can focus pulse in such a way to identically transform neighboring pairs of ions \cite{herold2016universal}. This allows one to perform the required single qubit rotations over the entire chain via sequential pairwise actions.

\begin{figure}[ht]
    \begin{center}
        \begin{tabular}{c}
            \includegraphics[width=.85\linewidth]{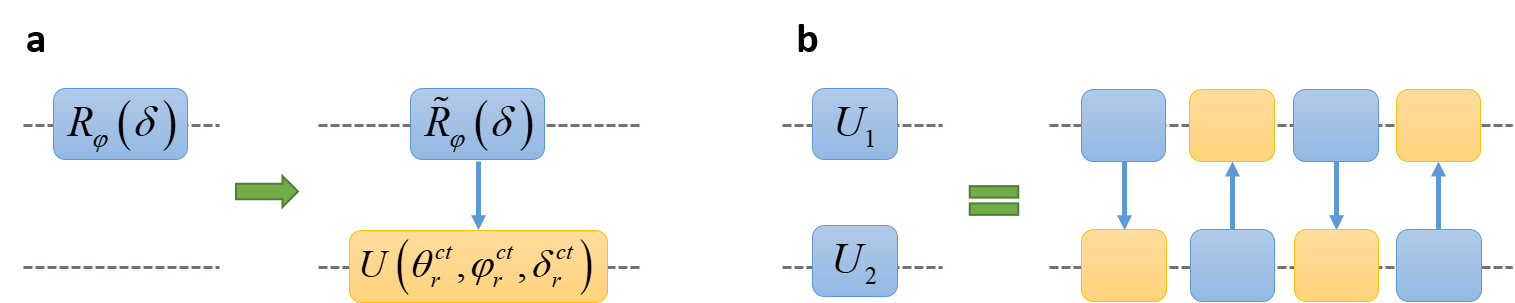}
        \end{tabular}
    \end{center}
    \caption{
        \label{fig:cross_talk}
        Single qubit gates cross-talk. (a) The transformation $R_\varphi(\delta)$ over the first qubit also rotates the second one with $U(\theta_r^{ct},\varphi_r^{ct},\delta_r^{ct})$. (b) To perform the required single qubit rotations it is sufficient to implement four non-ideal transformations.
    }
\end{figure}

Here we consider an individual addressing in the chain of two ions. The transformation of the first ion is non-ideal by itself (Section~\ref{sect:gates_angles}) and also affects the neighboring one transforming it with $U(\theta_r^{ct},\varphi_r^{ct},\delta_r^{ct})$ (Figure~\ref{fig:cross_talk}a). 

By comparing the parameters of the expected single qubit gate with its real effect on both qubits, one can find the optimal parameters for performing the required transformations. For errors within a few percents, the implementation of an arbitrary transformation $U_1$ over the first qubit and $U_2$ over the second one requires four non-ideal transformations (Figure~\ref{fig:cross_talk}b). In particular, one can set $U_2 = I$. A sequence of four gates then provides the required single qubit rotation over the first qubit, leaving the state of the second one unaffected.

Parameters of $U(\theta_r^{ct},\varphi_r^{ct},\delta_r^{ct})$ depending on $(\varphi,\delta)$ can be estimated by means of QT. As in Section~\ref{sect:gates_angles}, we define them using the linear model of the form \eqref{eq:linear_model}. The only difference here is that the transformation itself is performed on one qubit, and its effect is evaluated on the other. Note that before performing the procedure, it is necessary to build a QT model for each qubit separately using the methods described in Section \ref{sect:noisy_qt}.

\section{DISCUSSION}\label{sect:discussion}

In Section~\ref{sect:noisy_qt}, we discussed the two-step construction of the QT model from a set of measurements. At the first step, we estimate the qubit readout errors. At the second step, we implement a set of quantum circuits. These circuits consist of quantum gates that are used in QT for the input state preparation and for the measurement in a particular basis. The data from these circuits are used to reconstruct the parameters of these gates.

Such estimation of gate parameters through the implementation of their various sequences is similar to the gate set tomography (GST) approach \cite{blume2017demonstration,nielsen2020probing,dehollain2016optimization}. The latter, however, has a very wide ambiguity of the result obtained. This ambiguity cannot be overcome by implementing a larger number of quantum circuits.

Our model contains significantly fewer parameters than the GST model, and gives an unambiguous result. This is partly achieved by using the model of classical readout errors in the form \eqref{eq:fuzzy_readout}. We also consider a model of unitary single qubit gates, excluding the effect of decoherence. Finally, we use the fact that for trapped ion based systems, all single qubit transformations $R_\varphi(\delta)$ for different $\varphi$ are equivalent and determined by the relative phase $\varphi$ of the resonant pulse. This allows us to take $\varphi=0$ for the estimation of $\sqrt{X}$ gate parameters.

From here one can see that GST limitations are mainly because it considers the quantum system as a ``black box''. Taking some assumptions about the physical aspects of the quantum system into consideration can significantly reduce the ambiguity. Our model, however, does not consider the non-unitary errors of QT gates. This limits the accuracy of our approach. Therefore, further improvement of this model and accounting for decoherence can improve the approach.

In Section~\ref{sect:gates_opt}, we discussed the possibility of improving single qubit logical operations. In contrast to the optimal quantum control, here we focus on the possibility of obtaining the required ideal transformations, having imperfect ones at our disposal. For this, we introduce a linear model of the dependence of the true gate parameters on the expected one. We then estimate the parameters of this linear model from a set of QT experiments. This approach makes it possible to obtain logic gates close to ideal using imperfect ones. A similar methodology allows one to mitigate ion cross-talk when performing single qubit operations.

Note, however, that the present work was aimed at correcting logical errors. In real quantum systems, decoherence lead to non-unitary errors, the correction of which is also an important task.

\section{CONCLUSION}\label{sect:conclusion}

In this work, we analyzed the possibility of using the results of quantum tomography to improve the characteristics of transformations of a quantum register based on trapped ions. For this, we developed a quantum tomography model that takes into account instrumental errors in the quantum state preparation and measurement. Such a model is necessary in order to obtain consistent estimates in quantum tomography. We used the resulting model to detect and correct unitary errors of quantum gates in numerical experiments. To do this, we implemented sequences of non-ideal gates with pre-estimated errors, which resulted in the required ideal evolution. It is shown that the developed approach makes it possible mitigate not only the single qubit gates errors, but also the cross-talk in single-qubit gates.

\acknowledgments 
 
The research is supported by the Leading Research Center on Quantum Computing (Agreement No.~014/20) and by Theoretical Physics and Mathematics Advancement Foundation ``BASIS'' (Grant No.~20-1-1-34-1).

\bibliography{report} 
\bibliographystyle{spiebib} 

\end{document}